# Spatial Photonic Reservoir Computing based on Non-Linear Phase-to-Amplitude Conversion in Micro-Ring Resonators

**Charis Mesaritakis[1], Kostas Sozos[2], Dimitris Dermanis[1], Adonis Bogris[2]**
1. University of the Aegean, Dept. of Information and Communication Systems Engineering, Palama 2, Karlovassi, 83200-Samos-Greece. Author e-mail address: cmesar@aegean.gr

2. University of West Attica, Dept. of Informatics and Computer Engineering. Aghiou Spiridonos, 12243, Egaleo, Athens, Greece

**Abstract:** We present a photonic reservoir computing, relying on a non-linear phase-to-amplitude mapping process, able to classify in real-time multi-Gbaud time traces subject to transmission effects. This approach delivers an all-optical, low-power neuromorphic dispersion compensator.  © 2020 The Author(s)

## 1. Introduction

Reservoir Computing (RC) is a bio-inspired computational paradigm that has been designed to inherit the ability of recurrent neural network (RNNs) in tackling time-dependent tasks, whereas at the same time circumvent the cumbersome training associated with RNNs[1]. From an architectural perspective, RCs consist of a randomly inter-connected and untrained hidden layer followed by a linear trainable output layer. Training procedure is restricted in the output layer, thus well-established algorithms can be readily implemented. From a practical perspective, the inherent randomness of RC's nodes fits really well to hardware implementations, where manufacturing deviations will not hinder performance, but on the contrary will contribute to the richness of the RC's dynamics [1]. In this context, integrated photonics is a proliferating platform for RCs due to elevated processing speed and negligible power consumption. On the other hand, a long-sought feature in photonic RC, is the realization of an all-optical nonlinear activation function. This has been partially achieved either by exploiting active photonic devices [2], Kerr/thermo-optic based non-linearities, at the cost of elevated input power level [3]. An interesting alternative, is the use of photodiodes as the main nonlinear mechanism [1], nonetheless, this last approach dictates the processing of signals exclusively in the electrical domain. In this work, we present numerical results concerning an alternative, all-optical, non-linear activation function and a corresponding RC, that relies on the non-linear conversion of phase modulated inputs to amplitude modulated outputs in micro-ring-resonators (MRRs). This process is not power related, it can enable large scale RCs and it can offer a tunable activation function. The efficiency of the proposed scheme, is validated through numerical simulations targeting two tasks: time-trace classification and the mitigation of transmission effects for a 25Gbaud pulse-amplitude-modulated (PAM-4) signal. Results, confirm improved performance compared to delay-based RCs, and Bit-Error-Rate (BER) values below HD-FEC for 45Km transmission.

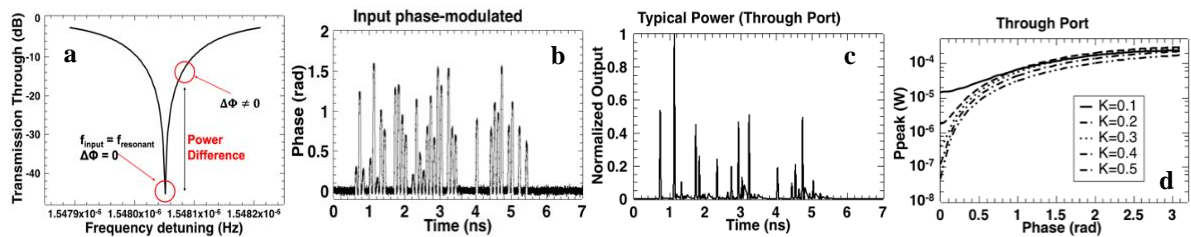

**Fig.1** a. Transfer function of an add/drop MRR for $k$=0.1, losses of 4.34 dB/m and R=100 μm. b) phase modulated input consisting of RZ pulses c) corresponding normalized optical power at the through port d) The peak power ($P_{peak}$) of the RZ pulse at the through port versus the phase variation at the input ($\Delta\varphi$) in rads.

## 2. MRR assisted non-linear phase-to-amplitude conversion

Typical MRRs can be designed either in an add/drop or notch configuration based on the number of straight waveguides coupled to the circular structure. In both cases, the optical power at the output(s) is governed by a phase matching condition between the optical field being transmitted through the straight waveguide ($A_{in}$) and the one circulating inside the MRR's cavity ($A_{MRR}$) [3]. In the typical case, where $A_{in}$ is amplitude modulated the phase matching condition is linked to $\varphi_{MRR}$, which is affected by the wavelength ($\lambda$), the refractive index ($n_{eff}$), the radius of the MRR ($R$) and $\tau$ and $k$ which are the transmission and coupling coefficient respectively (eq. 1-2).

$$A_{thr} = (1 - \tau)A_{in} - jkA_{MRR} \cdot e^{-j\varphi_{MRR}} \quad (1) \text{ and } \varphi_{MRR} = -\pi \cdot n_{eff} \cdot \left(\frac{2\pi}{\lambda}\right) \cdot R \quad (2)$$

In the case that the input signal is phase modulated, the difference between the two fields will be affected also by a term $\Delta\varphi(t)$, which is the instantaneous variation of the input signal's phase. The effect of this mismatch is equivalent to the variation of the input signal's frequency, that in turn results to a power variation at the MRR's output (fig.1a). We modelled a silicon-on-insulator add/drop MRR with $R=100$ μm, losses equal to 4.3 dB/m and symmetric coupling coefficient of $k=0.1$. In order to demonstrate the phase to amplitude mapping, we use input signals consisting of a phase modulated, return-to-zero (RZ) pulses with constant amplitude ($P=10$ μWatt) duration of 50 ps and 25 ps fall/rise time, whereas pulses' phase ranges from 0 to $\pi/2$ (fig.1b). In Fig.1c the corresponding optical power at the through port is presented. It can be seen that the input's phase transitions are imprinted at the power recorded at the MRR's output. In order to validate the non-linear relationship between input phase and output power, we varied the "depth" of the phase modulation from $\Delta\varphi=0$ to $\Delta\varphi=\pi$ assuming pulses as described above and we recorded the corresponding peak power variation at the output; we repeated this process for MRRs having different $k$. In Fig.1d it can be seen that by increasing $\Delta\varphi$, variation of the peak power follows the shape of the MRR's transfer function, resembling a saturation function. As coupling coefficient increases, the extinction ratio of the transfer function also increases thus the shape of the non-linear response varies. This effect can be triggered also by varying losses and radius. Interestingly, the drop port of the MRR provides also a frequency-detuning like response.

## 3. Photonic phase-to-amplitude MRR-RC

Aiming to exploit this process in a neuromorphic context, we expanded the single MRR numerical model so as to construct a multi-MRR RC. We assumed a simple loop, fabrication-friendly topology. In this configuration each RC's output (through port) is connected to the input port of the subsequent MRR, while the drop port in each node is fed to a photodiode with 10 GHz bandwidth (fig.2a). The intra-MRR connections (synapses) are assumed to be waveguides, with random transmission coefficient ranging $K_w=0.9-1$ and delay $\varphi_w=3$ ps-8 ps.

The first benchmark test devised, consists of generating 8 different input patterns. Each pattern comprises a series of 50, 10Gbit/sec phase-modulated pulses, as described above (fig.1b). The first 10 pulses in each pattern differ, while the rest 40 pulses are identical among patterns. Therefore, RC's memory is a critical parameter. Furthermore, aiming to generate a valid data set we add optical noise to each pattern by assuming the existence of an erbium doped fiber

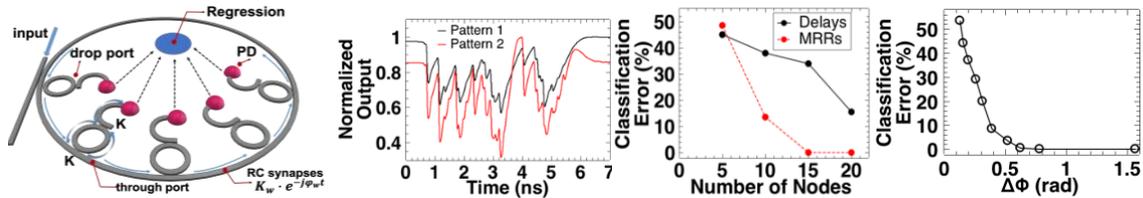

**Fig.2** a. the proposed RC, b) Normalized outputs for two different patterns. Classification error c) versus the number of nodes for delay-RC (black-solid) and for RC-MRR (red-dashed). d) versus phase modulation depth for 10-node RC with $k=0.3$ amplifier with variable gain at the RC's input. Through this approach, the input power is preserved constant but the signal-to-noise ratio (SNR) of the patterns can be adjusted. In fig.2b two characteristic electrical outputs recorded at the second drop port are presented that correspond to two different input patterns. It is evident that even though their last 40-pulse part is identical, the dimensionality expansion property of RCs is manifested. The labeled outputs were fed to a typical ridge regression scheme and classification error was computed. We first investigated the impact of the number of MRRs nodes in the RC's performance. As a basis for our investigation we assumed an RC-loop, where each MRR was replaced by a simple unity transfer function, thus eliminating all non-linear elements apart from the application of the photodiode at the output. Thus, this scheme mimics an optical delay line RC [2]. In fig.2c (solid-black) it can be seen that for the delay-RC, as the number of linear outputs increases (waveguides) classification error decreases slowly; for 20 nodes classification error equals to 15,6%. On the contrary if MRRs are included in the scheme, classification error exponentially decreases with the number of MRRs. When the MRR-RC has the same number of nodes ($n=20$) as the delay RC, error scales down to 0.037%. It is clear that the MRR induced non-linearity, significantly boosts performance. In the same context, we assumed a 15 MRR RC ($k=0.3$), and we varied the depth of phase modulation; meaning that we kept the minimum phase value equal to zero, while we varied the maximum value from 0 (no phase modulation) to $\pi$. By inspecting fig.1d it can be seen that this action affects node's non-linearity; small depth equals to an almost linear behavior. It can be seen that classification error is directly linked to the degree of non-linearity for $\Delta\varphi=0.13$ rads error is 53% whereas for $\Delta\varphi=\pi/2$ scales down practical to 0%.

The proposed scheme is benchmarked in a second "real-world" task comprising dispersion compensation of 25GBaud PAM-4 signal after transmission. We simulate the transmission, with the integration of Nonlinear Schrodinger equation using the Split-step Fourier method [4]. Then, we use the RC system so as to classify the distorted signals and reduce BER that builds-up as a result of chromatic dispersion. In the simulations, we consider all types of noises (thermal noise, shot noise) and analog-to-digital converters with 8-bit resolution. We generate 40000 pseudorandom PAM-4 symbols, drawn from a uniform distribution. The PAM-4 samples after transmission are used so modulate the phase of an optical carrier with frequency matching the resonance of the MRR. In this mapping process, the detected PAM-4 symbols are mapped to a phase space ranging from 0 to $2\pi$, so as to exploit the MRR's nonlinearity (see fig.2d). The RC's nodes were increased to 30 nodes and were arranged into 6 sub-RCs, each bearing 5 MRRs that receive the same input and deliver their outputs to the same output layer, so as to boost performance [5]. Furthermore, the *R* and *k* of all MRRs were perturbed from the nominal values of *k=0.3* and *R=100μm*, following a normal distribution (*σ=0.1*), so as to enrich the RC's dynamics. Training procedure comprises ridge regression using 50% of the symbols. In addition, regression's input was divided in batches of 10-25 symbols, thus providing the necessary memory for handling transmission induced inter-symbol-interference [6]. In fig.3, BER is evaluated versus transmission distance; mean-BER corresponds to the mean performance of 100 random RCs, while minimum-BER is linked to the optimum RC configuration. It can be easily seen that RC can optimally offer BER<$10^{-3}$ for transmission distance below 50 km, whereas for 40Km, which is a typical scenario for access and metro networks, BER is in the order of $4\cdot10^{-4}$. The RCs outperform linear classifiers for all transmission lengths. It is of utmost important to stress out that this performance could potentially come with very small power consumption during inference and without cumbersome training due to the nature of the RC.

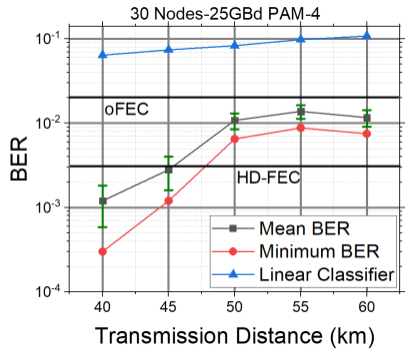

**Fig.3** BER versus transmission distance for a 25Gbaud/sec PAM-4 signal processed by a 30 MRR-RCs with random parameters (mean-black, squares) and optimum MRR-RC (red-circles). Performance for a linear classifier, an o-FEC and HD-FEC are included

**4. Conclusion**

In this work we demonstrated numerical results concerning an alternative approach to realize a non-linear activation function in integrated photonic neuromorphic systems. The proposed approach is based on the non-linear conversion of phase modulated signals to amplitude modulated outputs in resonant devices such as MRRs. This approach alleviates the need for active photonic neural nodes or for elevated optical power at the neural network's input. The form and magnitude of the nonlinearity can be shaped by properly choosing MRR parameters and ring configuration. Aiming to demonstrate the validity of our approach, we chose an RC architecture, which fits really well to the photonic platform and devised two scenarios: a typical time-trace classification task and amending the impairments due to transmission effects. Results demonstrate superior performance compared to delay-based RCs with negligible power consumption and system complexity.

**Acknowledgements**
This project has received funding from the EU H2020 NEOTERIC project (871330). K. Sozos is funded by the Hellenic Foundation for Research and Innovation (HFRI) and the General Secretariat for Research and Technology (GSRT), under grant agreement No 2247 (NEBULA project)